\documentstyle[aasms4]{article}

\def\taut{\tau_{\rm T}}

\def\msun{{\,M_\odot}}

\newcommand\tff{\tau_{\rm ff}}

\newcommand\fx{F_{\rm x}}

\def\refindent{\par\noindent\hangindent=3pc\hangafter=1 }

\def\aasup#1#2#3{\refindent#1, A\&AS, #2, #3}

\def\apj#1#2#3{\refindent#1, {\it ApJ}, {\bf#2}, #3.}
\def\apjlett#1#2#3{\refindent#1, {\it ApJL}, {\bf #2}, #3.}

\def\mnras#1#2#3{\refindent#1, {\it MNRAS}, {\bf#2}, #3.}
\def\nature#1#2#3{\refindent#1, {\it Nature}, {\bf #2}, #3.}

\def\>{$>$}
\def\<{$<$}

\def\simlt{\lower.5ex\hbox{$\; \buildrel < \over \sim \;$}}
\def\simgt{\lower.5ex\hbox{$\; \buildrel > \over \sim \;$}}
\def\sqr#1#2{{\vcenter{\hrule height.#2pt
      \hbox{\vrule width.#2pt height#1pt \kern#1pt
         \vrule width.#2pt}
      \hrule height.#2pt}}}

\begin{document}

\centerline{Submitted to the Astrophysical Journal {\it Letters}}

\bigskip
\title{X-rays From Magnetic Flares In Cygnus X-1: A Unified\\ Model
With Seyfert Galaxies} 
\author{Sergei Nayakshin$^{\dagger}$ and Fulvio Melia$^*$ 
\thanks{Presidential Young Investigator.}}
\affil{$^{\dagger}$Physics Department, The University of Arizona,
Tucson AZ 85721\\ $^*$Physics Department \& Steward Observatory, The
University of Arizona, Tucson AZ 85721}




\altaffiltext{1}{Presidential Young Investigator.}


\begin{abstract}
Some recent work, e.g., by Zdziarski et al., has shown that the spectrum of
Seyfert 1 Galaxies is very similar to that of several Galactic Black Hole 
Candidates (GBHCs) in their hard state.  However, several physical
constraints seem to rule out the two-phase model (otherwise successful
in the case of Seyfert Galaxies) for GBHCs.  Here, we show that this
conclusion is based on a number of key assumptions about the X-ray 
reflection/reprocessing within the cold disk that probably are not valid 
when the overlying corona is patchy, e.g., when it is comprised
of localized magnetic flares above the disk.  We find that if the
X-rays are emitted within these magnetic structures atop the cold disk, 
then the X-ray reflection/reprocessing of the primary X-rays by the latter 
is {\it not} scale-invariant with respect to the black hole mass. 
In particular, we show that in GBHCs the energy deposited by the
X-rays cannot be re-radiated fast enough to maintain equilibrium, unless
the X-ray skin heats up to the Compton temperature, at which point
the gas is mostly ionized.  This leads to a substantially reduced 
cooling rate for the active regions due to the correspondingly smaller
number of re-injected low-energy photons. We model this effect by 
introducing a transition layer situated between the corona and the cold 
disk, and find that the resulting spectrum is harder than that
obtained with the standard (and unrealistic) two-phase model. 
We apply this model to Cygnus X-1 and show that it {\it can} account for its 
observed spectrum.  This analysis therefore seems to provide a consistent
picture for both Seyfert Galaxies and GBHCs within the same framework,
with differences arising due to the changing physical conditions
in the two categories of sources, rather than due to an {\it ad hoc}
variation of the model parameters. 
\end{abstract}


\keywords{accretion disks --- black hole physics ---  Cygnus X-1 ---
galaxies: Seyfert --- magnetic fields  --- radiative transfer}


%

\section{Introduction}
The progress made in recent years in understanding the X-ray spectra
of Seyfert Galaxies and Galactic Black Hole Candidates (GBHCs)
indicates that the reflection and reprocessing of incident X-rays into
lower frequency radiation is an ubiquitous and important process.  For
Seyfert Galaxies, the X-ray spectral index hovers near a ``canonical
value'' ($\sim 0.95$; Pounds et al. 1990, Nandra \& Pounds 1994;
Zdziarski et al. 1996), after the reflection component has been
subtracted out of the observed spectrum.  It is generally believed
that the universality of this X-ray spectral index may be attributed
to the fact that the reprocessing of X-rays within the disk-corona of
the two-phase model leads to an electron cooling rate that is roughly
proportional to the heating rate inside the active regions (AR) where
the X-ray continuum originates (Haardt \& Maraschi 1991, 1993; Haardt,
Maraschi \& Ghisellini 1994; Svensson 1996). It has been suggested
that the ARs are probably magnetically dominated structures, i.e.,
magnetic flares (Haardt et al. 1994; see also Galeev, Rosner \& Vaiana
1979). 

GBHCs have similar, though considerably harder, spectra and a
reflection component less prominent than that of Seyfert Galaxies
(Zdziarski et al. 1996).  Indeed, Rossi X-ray observations of Cygnus
X-1 show no significant evidence for a reflection component (Dove et
al.  1997a).  It is the relatively harder spectrum and the small
reflection component (together with other more subtle arguments) that
led Dove et al. (1997a,b), Gierlinski et al. (1997) and Poutanen,
Krolik \& Ryde (1997) to conclude that the two-phase patchy corona model
does not apply to Cygnus X-1 -- arguably the best studied GBHC.
Physically, the harder spectrum of GBHCs requires a much lower cooling
rate than can be expected in the two-phase model. However, as we shall
show below, this conclusion relies heavily on the assumption that the
reflection/reprocessing process in the GBHCs is similar in nature to
that of the higher mass Seyfert nuclei.  In view of the fact that in
most situations the accretion disk physics is black hole
mass-invariant, this at first appears to be a reasonable assumption
(but see Ross, Fabian \& Brandt 1996).

In this {\it Letter}, we point out that previous studies of the X-ray 
reflection process (e.g., Sincell \& Krolik 1997;  Magdziarz \& Zdziarski 1995; and
additional references cited in Nayakshin \& Melia 1997a) have concentrated on 
a static, full corona, a model that is unlikely to work even for Seyfert
Galaxies (e.g., Haardt et al. 1994; Svensson 1996).
Motivated by this fact, Nayakshin \& Melia (1997a; hereafter paper I) 
investigated the X-ray reflection process in AGNs assuming that the ARs are 
magnetic flares above the disk. These flares are short lived 
(see below), and the properties of the {\it transient} X-ray reflection 
turn out to be similar, and yet different enough from the static reflection 
process, to provide an interesting explanation for the narrow range in 
the observed temperature of the Big Blue Bump and the low degree of
ionization in the underlying disk.

It is therefore important to consider the physics of reflection in the 
two-phase model for GBHCs with transient ARs, paying particular attention
to aspects of this process that are not scale-invariant with respect to the 
black hole mass $M$.  Below we show that a transient X-ray irradiation of the disk 
may account for the differences in the reflected fraction and the overall 
spectrum of Seyfert Galaxies and GBHCs. We find that the X-ray skin in GBHCs 
is much hotter and more strongly ionized than that in Seyfert nuclei, and 
is probably at the local (optical depth dependent) Compton temperature. 
It therefore acts as a ``transition'' layer in the case of GBHCs.
This is in contrast to the situation in static X-ray reflection models, 
where the pressure equilibrium condition in this skin is black hole 
mass-invariant and leads to a cooler and far less ionized plasma.

As we shall see, the reflection process in a transition layer held at the 
Compton temperature substantially reduces the Compton cooling rate within
the corona due to the upscattering of reprocessed radiation.
We present several test results
indicating that the presence of this transition layer, heated and cooled 
by Compton scattering with the hot coronal and cold disk radiation
leads to a spectrum that is much harder than that of the standard 
two-phase model.  Our results demonstrate that a consideration of the 
structure of the X-ray skin in a realistic two-phase model may
therefore remove many current inconsistencies of the coronal picture
with observations. We point out that the two-phase model with 
magnetic flares as the particle energizing mechanism in active 
regions may constitute a single explanation for both Seyfert nuclei
and GBHCs, and account for differences in their spectra self-consistently.

\section{Transient X-ray Reflection For GBHCs}

As shown in paper I, the main characteristic of transient reflection,
distinguishing it from its well-studied static counterpart, is that
magnetic flares can only be active during a disk hydrostatic time
scale.  It is not difficult to show (e.g., using the model of Svensson
\& Zdziarski 1994) that the photon diffusion time across the disk is
much longer than this for both radiation and gas-dominated
disks. Therefore, no thermal equilibrium can be established between
the underlying cold disk and the incident X-radiation during the
flare. Instead, a quasi-equilibrium is established within an X-ray
skin with Thomson optical depth $\taut\sim$ few.  It is this X-ray
skin that plays the major role in reprocessing and reflecting the
incident X-rays. We should further refine our definition of the
X-ray skin (or transition layer) by specifying that this layer
corresponds only to that part of the disk's upper region below or
close to a magnetic flare, i.e., within a few scale heights
of the AR, since that is where most of the X-ray reprocessing will 
occur. Thus, the X-ray skin is a transient phenomenon: it comes 
into existence due to the illuminating X-ray flux from the AR, 
so it lasts only as long as the flare persists. 

Under the very intense incident X-ray flux, the upper layer of the
disk in Seyfert nuclei contracts to a high density, which allows the
irradiated gas to emit the deposited energy efficiently due to the
strong density dependence of the optically thin free-free process
(e.g., Rybicki \& Lightman 1979).  However, as was noted in paper I,
this treatment is valid only when the X-ray skin is optically thin to
free-free emission.  Using a Rosseland mean free-free optical depth
and Equation (4) of paper I, it is straightforward
to see that the free-free optical depth of the X-ray skin is $\tff
\simeq 5\times 10^{-5} (\taut/3)^2\, l_2^2 \,\Delta R_{13}^{-1}
T_6^{-7/2}$, where $l \equiv 100\; l_2 \sim 100$ is the compactness
parameter of the AR, $T_6$ is the electron temperature in units of
$10^6$ K, and $\Delta R_{13} \equiv \Delta R/10^{13}$ cm is a
typical size of the AR, expected to be of the order of the disk scale
height. In Seyfert nuclei, $\tff$ is very much smaller than unity,
validating the optically thin approximation.  However, rescaling the
parameters for the case of GBHCs with a mass $\sim 10 \msun$, we
obtain
\begin{equation}
\tff \simeq  500 \times (\taut/3)^2 {l_2^2\over \Delta R_{6}}
T_6^{-7/2}\;.
\end{equation}

The most straightforward way to modify this result would be to simply
assume blackbody emission, but this approach would be incorrect since
the X-ray skin cannot consistently be optically thick and in pressure
equilibrium, as we now will show. The latter is set up between the
incident X-ray (ram) pressure and the internal skin pressure, with
contributions from both the gas and the reprocessed radiation, which
was neglected in paper I.  In Seyfert nuclei the skin's internal 
gas and radiation pressures are comparable, but in GBHCs the
radiation pressure would be dominant since a typical photon suffers
many absorption/emission events before escaping from the gas.  In 
equilibrium, the internal radiation pressure must be smaller than 
the external X-ray pressure, so that
\begin{equation}
\tau F_{\rm bb}/c \lesssim \fx/c\;,
\end{equation}
where $\fx$ is the incident X-ray flux from the AR, $\tau$ is the
total optical depth of the X-ray skin, and $F_{\rm bb}$ is the 
blackbody flux out of the transition layer. On the other hand, if the
incident energy is reradiated out of the transition layer, one should
expect that
\begin{equation}
2 F_{\rm bb} \sim \fx\;,
\end{equation}
where the factor $2$ arises because the radiation can escape in both
the upward and downward (i.e., toward the disk midplane) directions.
Thus, unless $\tau \lesssim$ few, a self-consistent equilibrium is not possible.

So what does happen in the irradiated layer in physical terms?  The
incident X-rays, being too energetic (most of the flux in Cygnus X-1
is at $> 20$ keV), are not absorbed via free-free absorption if the
gas temperature is $T \ll$ few keV.  Under these conditions, the
X-rays would penetrate unimpeded into the skin to a Thomson optical
depth $\sim$ few, just as in the static case. Accordingly,
irrespective of how large $\tff$ is, the incident X-rays would heat
the upper region of the disk with $\taut <$ few.  However, the
down-scattered X-rays, and the UV to soft X-ray photons due to
internal emission would be readily absorbed and could not escape fast
enough from the X-ray layer.  Therefore, we have a situation where the
energy penetrates into the skin easily, but cannot leave due to
multiple absorption/emission events (one should recall that blackbody radiation
is appropriate only when an average photon scatters many times before
it escapes), and thus no equilibrium between heating and cooling can
be established. As the reprocessed radiation builds up inside the
X-ray skin, it reaches the point where its pressure exceeds the
external X-ray ram pressure, and the X-ray skin expands, becoming
hotter and less dense. Correspondingly, the free-free emissivity
drops, and the X-ray skin cannot keep up with the energy deposition
rate. The heating process itself, however, has its limitations: at the
local Compton temperature (i.e., about $15$ keV) the radiation field
does not transfer any net energy to the gas by inverse Compton
interactions. The gas simply {\it reflects} most of the incident
X-rays, rather than absorbing and reprocessing them into the UV range.

This contrasts with the static case, in which the nature of the
pressure equilibrium within the X-ray skin is quite different (Sincell
\& Krolik 1997).  Here, the transient illuminating X-ray flux $\fx$ is
much larger than in the static case, because the same fluence (i.e.,
integrated flux) must now be produced from a smaller coronal region
and, in addition, the ARs are only active for a short time.  Taken
together, these lead to a larger ionization parameter $\xi_{\rm x}
\simeq 2.5 \times 10^{3} (\fx/c P_{\rm gas})$, where $P_{\rm gas}$ is
the gas pressure within the X-ray skin (see, e.g., Eq. 3.1 of Zycki et
al. 1994).  Since $P_{\rm gas}$ is certainly smaller than $\fx/c$,
$\xi_{\rm x}$ is very large, and the GBHC X-ray skin is therefore in
the ``limit of a hot medium'' (see \S 3.1 of Zycki et al. 1994).  The
fact that this segment of parameter space is unlikely to be reached in
a static corona is probably the reason why the spectroscopic
consequences of reflection in a Compton equilibrium medium have not
been considered earlier.  We take up this question in the following
section.

\section{``Three-Phase'' Model}

Hereafter, we shall assume that the X-ray skin attains its {\it local}
Compton temperature. We conduct several representative tests which
will allow us to understand the important physics of the model,
without necessarily attempting yet to fit any particular spectrum of a
GBHC. Our model consists of an active region above the (heated)
transition layer. The geometry of the AR is probably closer to a
hemisphere than a slab, but for simplicity we shall adopt the latter
for the radiation transport, neglecting the boundary effects.  However
crude, this approximation is adequate for our purposes.  Experience
has shown that spectra produced by Comptonization in different
geometries are usually qualitatively similar (i.e., a power-law plus
an exponential roll-over), and it is actually the fraction of soft
photons entering the corona that accounts for most of the differences
in the various models, because it is this fraction that affects the AR
energy balance. 

Following the standard practice in ionization/reflection calculations,
we model the reflecting medium as being one dimensional, with its only
dimension being the optical depth into the disk (measured from the
top).  Nevertheless, we have to take into account the fact that the
patchy corona geometry permits some part of the reprocessed radiation
to re-enter the corona, whereas the rest escapes to the observer
directly. Accordingly, the observed spectrum consists of the direct
component, emerging through the top of the corona (AR), and a fraction
$\Omega$ of the reflected radiation that emerges from the transition
layer and does not pass through the corona on its way to us.  A
fraction $g = 0.5$ of the reflected spectrum goes back into the corona
through its bottom (cf. Poutanen \& Svensson 1996). Below the
transition layer lies an optically thick portion of the disk with
Thomson optical depth $\taut\gg 1$, held at a temperature $T_{\rm bb}
= 100$ eV. We employ the Eddington (two-stream) approximation for the
radiative transfer in both the AR and the transition layer, using both
the zero (isotropic) and first order moments of the exact
Klein-Nishina scattering kernel (Nagirner \& Poutanen 1994). The
optical depth of both the transition layer $\tau_{\rm trans}$ and the
corona $\tau_{\rm c}$ are treated as parameters; $\tau_{\rm c}$ is
fixed at an arbitrarily chosen value of $0.7$ for the purpose of
demonstrating the main point.  In a generally accepted setup for the
two-phase model, the transition layer is absent, and the X-rays
incident on the cold disk below the AR are partially reflected
($10-20\;\%$), while the rest are reprocessed and re-radiated as
blackbody radiation. In the tests reported here, the incident X-rays
are first scattered within the transition layer, as we self-consistently
calculate the spectrum that is incident on the cold disk after the
original spectrum from the AR passes through the layer. The spectrum
incident on the cold disk from the transition layer is reflected and
reprocessed in the standard manner (Magdziarz \& Zdziarski 1995), and
then re-enters the transition layer from below. We assume a coronal
heating rate much exceeding the local intrinsic disk flux and find the
radiation field and the self-consistent temperature in both the corona
and in the transition layer as a function of the optical depth.

Figure 1 shows the ``observed'' spectrum for several values of
$\tau_{\rm trans}$: $0$, $0.6$, $2.5$, and $10$, with $\Omega=0.5$.
It can be seen that the spectrum hardens as $\tau_{\rm trans}$ increases. 
To help explain why this happens, we plot
in Figure 2 the integrated albedo $a$ for photons with energy $E> 1$
keV as a function of $\tau_{\rm trans}$. The albedo is simply the
inverse ratio of the incident flux in the given energy range to the
returning one, i.e., that emerging from the top of the transition
layer. As $\tau_{\rm trans}$ increases, a large
fraction of the photons from the AR are reflected before they have a
chance to penetrate into the cold disk where the blackbody component
is created. Therefore, a smaller flux of energy is deposited below the
transition layer, which leads to a decreased cooling from the
Comptonization of soft, reprocessed radiation.  For a moderate optical
depth $\tau_{\rm trans}$, this result is quite insensitive to the
temperature in the transition layer as long as Fe is highly ionized.
We checked this by simply setting the transition temperature at the
arbitrarily chosen values of $1.5$ and $6$ keV, instead of the
self-consistent temperature distribution calculated above, which
varied (with optical depth into the transition layer) from about $2$
to $4$ keV for the respective values of $\tau_{\rm trans}$.  We found
that the relative variations in the spectrum and the albedo resulting
from this were less than about $3\;\%$.  For higher optical depths
($\tau_{\rm trans}\gtrsim 4$), pre-Comptonization of the soft disk
radiation becomes important and additionally decreases the Compton
cooling of the corona by this component, so that the temperature of
the transition layer becomes essential.

Figure 3 shows the observed spectrum (solid curve), comprising the
intrinsic AR spectrum (short-dash) and the reflected component
(emerging from the top of the transition layer; dotted curve)
multiplied by $\Omega = 0.5$. Also shown is the reprocessed component
at the base of the transition layer (long-dash).  All the intensities
propagate in the upward direction. Notice that due to the presence of
the transition layer, the reflected component is much harder than the
reprocessed component, which would be the ``normal''
reflection/reprocessing component without this layer. Notice also that
the bump around $\sim 40$ keV normally attributed to the reflected
component is broad (compared with the long-dashed curve), and so the
reflected component is here less noticeable.  

Furthermore, the combined power below $2$ keV accounts for only
$25\;\%$ of the total, whereas the corresponding fraction is about
$50\;\%$ in the standard (static) two-phase model. This large power
coming out in low energy photons was the main reason why the standard
two-phase corona-disk model failed to account for the observations.
The large soft photon power is not only not observed in the
Cygnus X-1 spectrum, it also provides too much cooling for the Active
Regions, and their spectra are never hard enough. In the
time-dependent situation these problems are resolved because of the
structure of the transition layer, as we described above.

The transition layer may be thought of as a partially transparent 
mirror. Crudely, some of the photons are scattered back without a 
change in energy, and the rest proceed to the cold disk and suffer 
the usual transformation into soft disk photons. Photons 
downscattered in the transition layer before they return to the 
corona do not contribute to the cooling because the energy gained by an
electron in the layer is later used to upscatter the softer photons
coming from the cold disk (whereas in the cold disk case this energy
would be used to produce the soft radiation). 

The spectral calculations reported here
could also be appropriate for the static patchy corona model if the
upper layer of the disk was hotter than usually assumed. Indeed, it is
not hard to imagine that the upper layer is being heated in a way
similar to heating of the Solar corona. If the temperature of the
layer is few keV and its Thomson optical depth is close to $\sim 3$,
then the spectra produced in this situation may be close to the
observed hard spectra of the GBHCs. It is not clear to us why this
possibility has never been explored by previous workers. At the same
time, even if such a static model could remove the problems for the
GBHCs spectra, one would need to explain why the upper layer of the
disk in Seyfert Galaxies is not being heated to similar
temperatures. Thus, the real strength of the magnetic flare model of
the active regions is in the fact that this is the same physics that
explains spectra of both Seyfert Galaxies and GBHCs.

\section{Discussion}

By considering the equilibrium structure of the irradiated
X-ray skin close to an active magnetic flare above a cold accretion
disk, we have shown that neither the optically thin free-free emission
process or the black body radiation can consistently comply with
pressure equilibrium constraints in the X-irradiated skin of GBHCs.
Thus, in sharp contrast to the transient X-ray reflection process in
Seyfert nuclei (paper I), the X-ray skin achieves {\it Compton equilibrium}
($k T\sim$ few keV) with the incident X-radiation.
As a result, most of the incident X-rays are Compton reflected
back into the AR before reaching the cooler disk material where
reprocessing into a soft-excess component occurs. Accordingly, the
amount of cooling due to the soft radiation re-entering the AR is
drastically reduced and this allows the two-phase model with magnetic
flares to reach the parameter space necessary to explain the observed
X-ray spectrum of Cygnus X-1.

The consequences of this ``three-phase'' structure include the
following: (1) GBHC spectra should be harder than those in typical
Seyfert 1s.  For $\tau_{\rm trans}\gg 1$, the spectrum may be somewhat
different from that of single cloud Comptonization plus a reflection
component, especially in the region of $\sim 30 - 200$ keV, which is
not inconsistent with data (see Fig. 3 and Gierlinski et al. 1997).
(2) The observed soft X-ray excess should contain comparatively less
power than the hard component, in contrast to Seyfert 1s.  (3) The
Thomson optical depth of the flares should be similar in GBHCs and
Seyfert Galaxies, and therefore so should the electron temperature
within their ARs (see Nayakshin \& Melia 1997b; this aspect of the
model does not depend on $M$).  (4) No anisotropy break should be seen
in the spectrum due to the input ``soft'' radiation not being a
blackbody and entering the ARs sideways (in accordance with
observations; see, e.g., Gierlinski et al. 1997). As was discussed in
Poutanen \& Svensson (1996), the anisotropy break occurs where the
second order scattering peaks. However, as we found from our numerical
results, the anisotropy break disappears as the optical depth of the
transition layer increases. The reason for this is physically
transparent: the reflected continuum is no longer the blackbody
emission (which was assumed by Gierlinski et al. 1997, and Poutanen \&
Svensson 1996) and is quite broad. Compton scattering broadens any
initial photon distribution, and therefore the second order scattering
of the reflected continuum becomes a very diffuse function, with a
shallow peak.  Among other effects that should reduce the anisotropy
break is the stratified temperature distribution below the AR and its
dependence with distance from the flare.  The cold disk emission will
then be a sum of black bodies with different temperatures, and will be
even broader than what we assumed in our calculations. Earlier
contrasting results found by Gierlinski et al. (1997) may be due to an
oversimplification of the cold disk structure and emission.  (5) The
reflected component in the observed spectrum must be less pronounced
or not observable, depending on the transition layer optical
depth. (6) The Fe lines from the inner accretion disk should be either
weak or broad and thus indistinguishable from the continuum due to
Comptonization within the transition layer. The observed weak
K$\alpha$ line may then be arising from the cold outer disk. (7) The
same is true for the Fe edge. Note that observationally, it is very
hard to detect a broad Fe edge in the case of Cygnus X-1 (Ebisawa
1997, private communication).  (8) X-ray variability should be
observed on a time scale of the order of the disk hydrostatic time
scale, i.e., $\sim$ milliseconds. Since one separate flare lives only
a few milliseconds, the hard and soft X-rays should vary
instantaneously down to a fraction of this, i.e. 1 millisecond or
so. Therefore, the observed high coherence of the hard and soft X-rays
in GBHCs (e.g., Vaughan \& Nowak 1997) is a natural consequence of the
model.

Although a more detailed spectral modeling is needed to confirm many
of these predictions for the hard state spectrum in GBHCs, even the
simple treatment used here shows that the two-phase model with
magnetic flares as ARs may ultimately account for both the
characteristics of Seyfert Galaxies and GBHCs. Very importantly, it
explains the differences in the spectra of these two classes of
objects self-consistently, i.e., based solely on the physics of the
irradiated region at the surface of the disk, rather than as a result
of an {\it ad hoc} variation of the parameters.

\section{Acknowledgments}

This work was supported in part by NASA grant NAG 5-3075.

%
%
%

{}

\newpage
\figcaption[]{The ``observed'' spectrum for various values (marked) of
the transition layer optical depth, $\tau_{\rm trans}$.}

\figcaption[]{Integrated albedo (reflected fraction) as a function of
the transition layer optical depth, $\tau_{\rm trans}$, for photons with energy
$> 1$ keV.  Also plotted (dotted curve) is the ratio of the observed hard
luminosity (above $2$ keV) to the observed total luminosity. Note that the
transition layer reflects a greater fraction of the photons as 
$\tau_{\rm trans}$ increases, and the spectrum correspondingly hardens.}

\figcaption[]{Decomposition of the observed spectrum (solid curve) into
its essential components: the intrinsic AR spectrum (short-dash) plus the 
reflected component (emerging from the top of the transition layer; dotted curve) 
multiplied by $\Omega = 0.5$. The reprocessed component at the 
base of the transition layer is also shown by the long-dashed curve. 
See text for a further discussion.}


\begin{thebibliography}{}

\bibitem[]{} Dove, J.B. et al. 1997a, submitted to ApJL, astro-ph/9707322

\bibitem[]{} Dove, J.B. at al. 1997b, astro-ph/9705130

\bibitem[Galeev, Rosner \& Vaiana 1979] {galeev} \apj{
Galeev, A. A., Rosner, R., \& Vaiana, G. S., 1979 }{ 229}{ 318}

\bibitem[]{} \mnras{Gierlinski, M. et al. 1997}{288}{958}

\bibitem[Haardt \& Maraschi 1991] {hm91} \apj{Haardt F. \& Maraschi L.
1991} {380}{ L51}
\bibitem[Haardt \& Maraschi 1993] {hm93} \apj{Haardt F. \& Maraschi L.
1993}{413}{ 507}
\bibitem[Haardt et al. 1994] {haardt94} \apj{Haardt, F., Maraschi, L.,
\& Ghisellini, G. 1994}{432}{ L95}

\bibitem[]{}\mnras{Magdziarz, P. \& Zdziarski, A.A. 1995}{273}{837}

\bibitem[]{np94a} Nagirner, D.I., \& Poutanen, J. 1994, Astrophysics
and Space Physics Reviews, v. 9

\bibitem[] {np94} \mnras{Nandra, K. \& Pounds, K. 1994}{268}{405}

\bibitem[Nayakshin \& Melia 1997a]{} \apjlett{Nayakshin, S. \& 
Melia, F. 1997a}{484}{L103}

\bibitem[Nayakshin \& Melia 1997b]{} \apjlett{Nayakshin, S. \& 
Melia, F. 1997b}{490}{L13}

\bibitem[] {p90} \nature{Pounds, K.A. et al. 1990}{344}{132}

\bibitem[]{} Poutanen, J, Krolik, J.H, \& Ryde, F. 1997, in the Proceedings of
the 4th Compton Symposium, astro-ph/9707244

\bibitem[]{} \mnras{Poutanen, J., Nagendra, K.N., \& Svensson, R. 1996}
{283}{892}

\bibitem[]{} \apj{Poutanen, J. \& Svensson, R. 1996}{470}{249}

\bibitem[]{} \mnras{Ross, R.R., Fabian, A.C., Brandt, W.N. 1996}{278}{1082}

\bibitem[Rybicki \& Lightman 1979] {ryb79} Rybicki, G. B. \&
Lightman, A.P., 1979, Radiative Processes in Astrophysics, John Wiley
and Sons: New York.

\bibitem[Sincell and Krolik 1997] {sk97} \apj{Sincell, M.W. \&
Krolik, J.H. 1997}{476}{605S}

\bibitem[Svensson \& Zdziarski 1994] {sz94}\apj{Svensson, R. \&
Zdziarski, A. 1994}{ 436}{ 599}

\bibitem[Svensson 1996] {sv96} \aasup{Svensson, R. 1996}{120}{475}

\bibitem[]{} \apjl{Vaughan, B.A. \& Nowak, M. A. 1997}{474}{L43}

\bibitem[]{zd96} \aasup{Zdziarski, A.A. et al. 1996}{120}{553}

\bibitem[]{} \apj{Zycki, P.T. et. al. 1994}{437}{597}


\end{thebibliography}
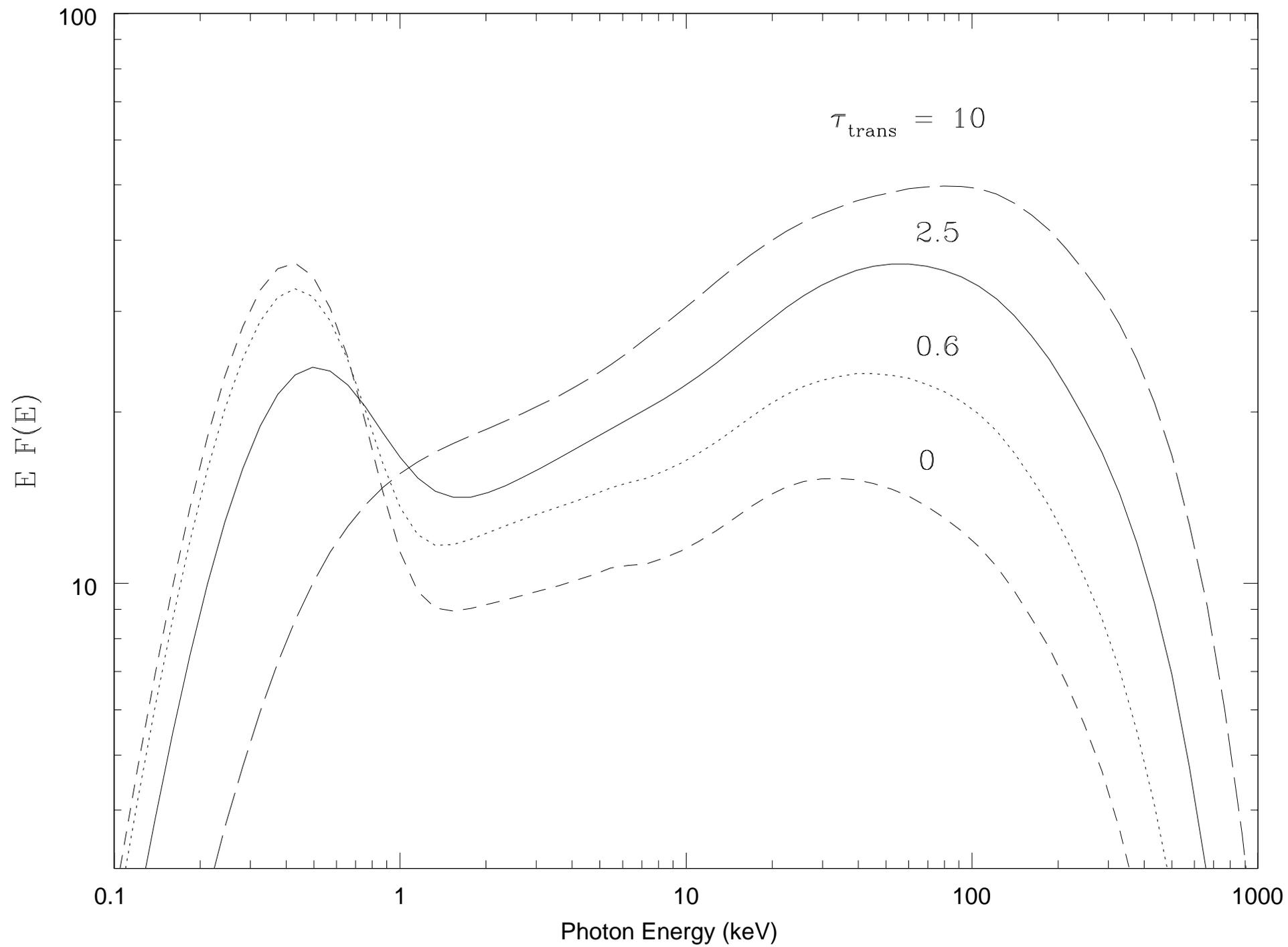
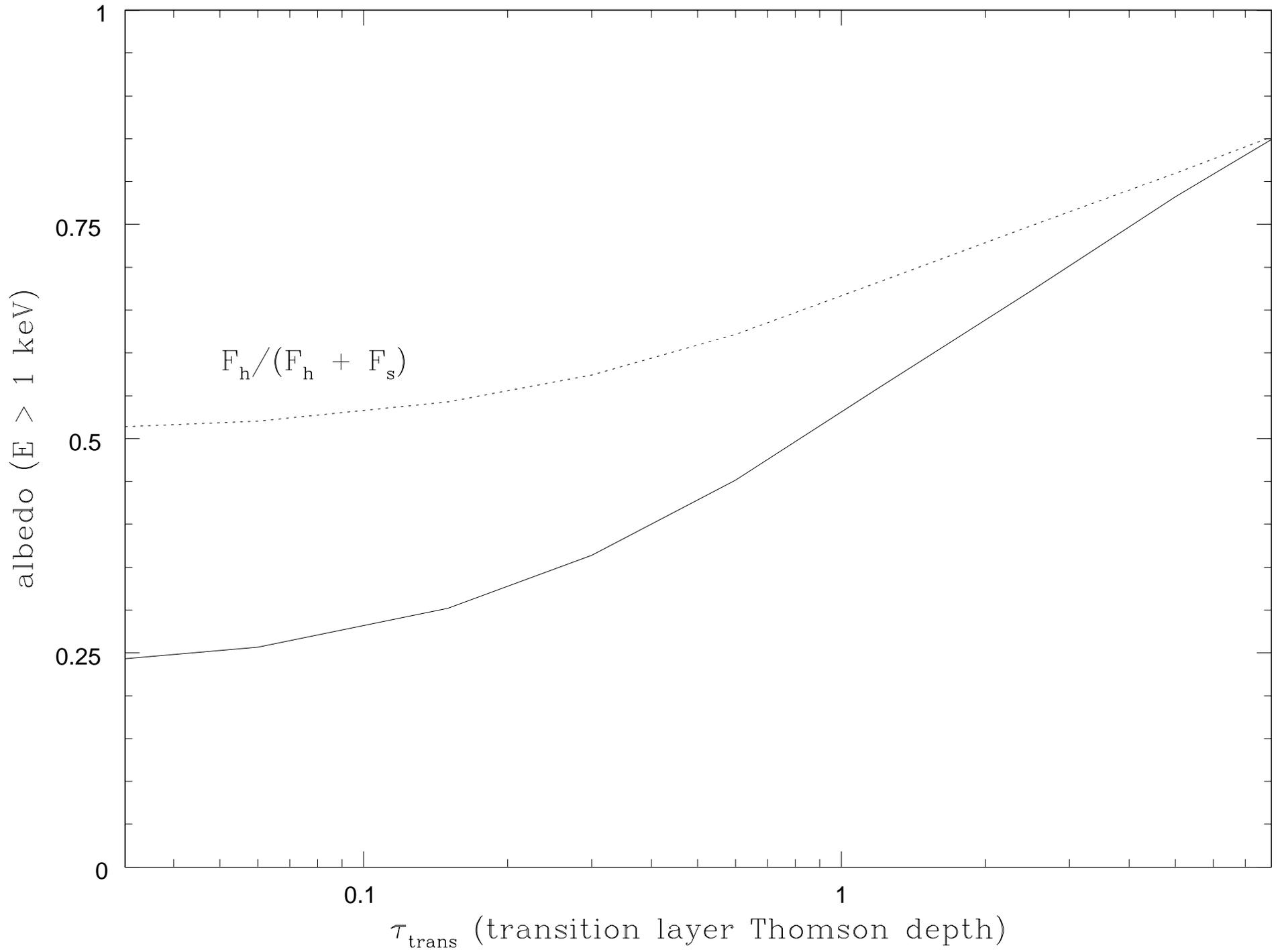
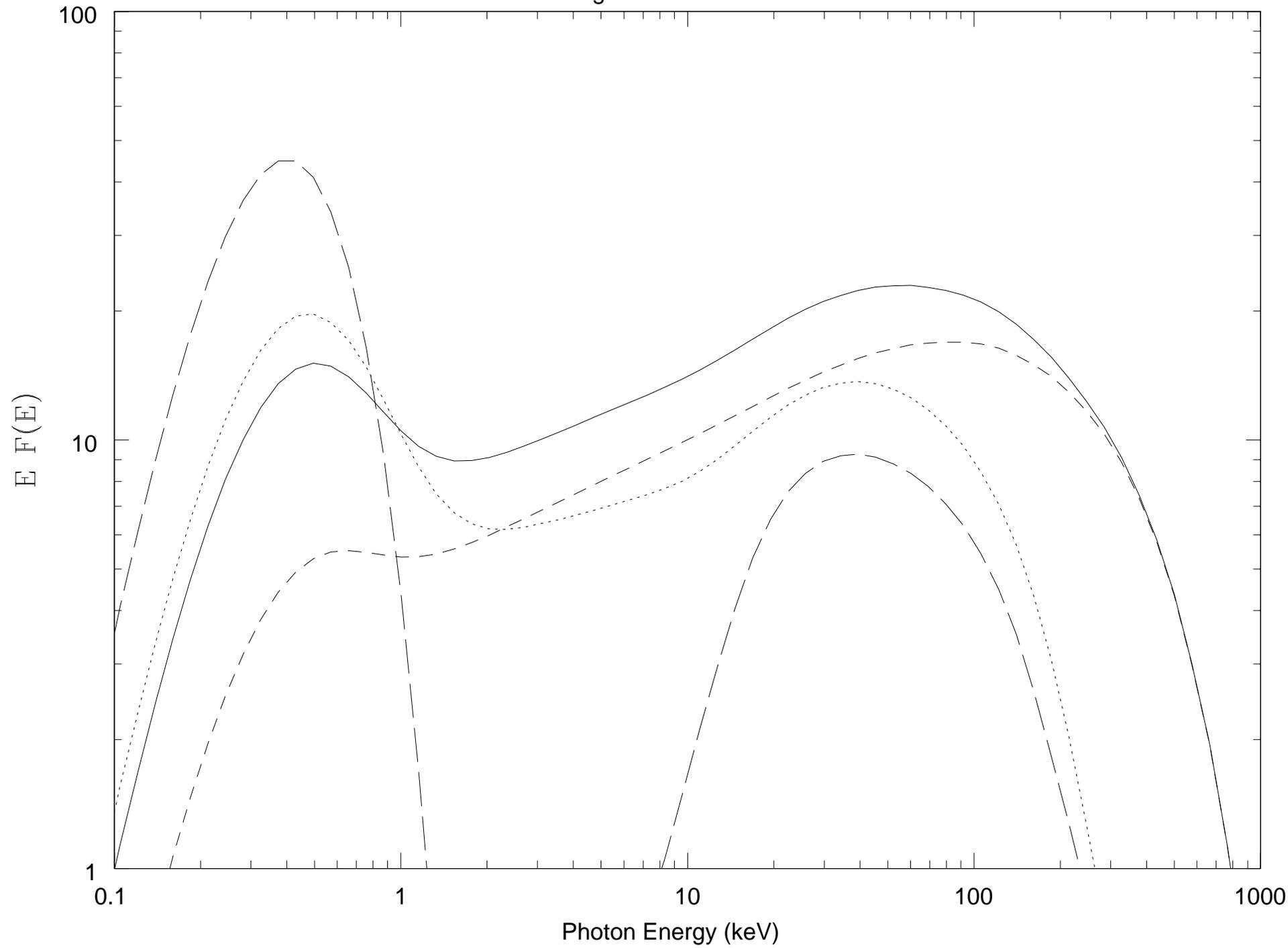
\end{document}